\documentclass[12pt,a4paper]{article}
\usepackage{a4wide}
\usepackage{latexsym}
\usepackage{amssymb}
\usepackage{graphicx}
\usepackage{amsmath, cite}

 \makeatletter
\@addtoreset{equation}{section} \makeatother

\def\bfone{\relax{\rm 1\kern-.35em 1}}

 \makeatletter
\@addtoreset{equation}{section} \makeatother

\def\be {\begin{equation}}
\def\ee {\end{equation}}
\def\bea {\begin{eqnarray}}
\def\eea {\end{eqnarray}}
\def\bc {\begin{center}}
\def\ec {\end{center}}

\def\D  {\Delta}

\def\bfg {\begin{figure}}
\def\efg {\end{figure}}
\def\bi {\begin{itemize}}
\def\ei {\end{itemize}}
\def\nn {\nonumber}

\def\la {\label}
\def\le {\left}
\def\ri {\right}

\DeclareFontFamily{U}{rsf}{} \DeclareFontShape{U}{rsf}{m}{n}{
  <5> <6> rsfs5 <7> <8> <9> rsfs7 <10-> rsfs10}{}
\DeclareMathAlphabet\Scr{U}{rsf}{m}{n}

\begin{document}

\begin{center}
{\bf \large{Minimal Length, Friedmann Equations and Maximum Density\\[10mm]}}
\large Adel Awad$^{
\dagger \ddagger}$\footnote{\upshape{Electronic address~:~\upshape{adel.awad@bue.edu.eg }}} and  Ahmed Farag Ali$^{\star \S}$\footnote{Electronic address~:~\upshape{ahmed.ali@fsc.bu.edu.eg,~~afarag@zewailcity.edu.eg }}
\\[7mm]
\small{$^\dagger$Center for Theoretical Physics, British University of Egypt}\\ \small{Sherouk City 11837, P.O. Box 43, Egypt.}\\[4mm]
\small{$^\star$Centre for Fundamental Physics, Zewail City of Science and Technology}\\ \small{ Sheikh Zayed, 12588, Giza, Egypt.}\\[4mm]
\small{$^{\ddagger}$Department of Physics, Faculty of Science, Ain Shams University, Cairo, 11566, Egypt.}\\[4mm]
\small{$^\S$Department of Physics, Faculty of Science, Benha University, Benha, 13518, Egypt}
\end{center}

\begin{abstract}
Inspired by Jacobson's thermodynamic approach\cite{Jacobson}, Cai et al \cite{Cai,Akbar:2006kj} have shown the emergence of Friedmann equations from the first law of thermodynamics. We extend Akbar--Cai derivation\cite{Akbar:2006kj} of Friedmann equations to accommodate a general entropy-area law.
Studying the resulted Friedmann equations using a specific entropy-area law, which is motivated by the generalized uncertainty principle (GUP), reveals the existence of a maximum energy density closed to Planck density. Allowing for a general continuous pressure $p(\rho,a)$ leads to bounded curvature invariants and a general nonsingular evolution. In this case, the maximum energy density is reached in a finite time and there is no cosmological evolution beyond this point which leaves the big bang singularity inaccessible from a spacetime prospective. The existence of maximum energy density and a general nonsingular evolution is independent of the equation of state and the spacial curvature $k$. As an example we study the evolution of the equation of state $p=\omega \rho$ through its phase-space diagram to show the existence of a maximum energy which is reachable in a finite time.

\end{abstract}


\section{Introduction}

A genuine connection between the laws of black holes mechanics and those of
thermodynamics is widely believed today. It started out with a mere analogy connecting
these two sets of laws which was first pointed out by Bardeen, Carter and Hawking\cite{Bardeen}. Only after the discovery of Hawking radiation \cite{Hawking}, it has been realized that these thermodynamic relations describe the thermal properties of a black hole. In this work Hawking was able to show that a black hole behaves quantum mechanically as a black body radiator, with a temperature proportional to its surface gravity and entropy proportional to its horizon
area \cite{Bekenstein}.

Based on the above connection between entropy and horizon area, and that of temperature and surface
gravity, Jacobson \cite{Jacobson} found that Einstein field
equations can follow exactly from the fundamental thermodynamic
relation between heat, entropy, and temperature $dQ=T dS$, i.e.,
Clausius relation. The argument is based on demanding that Clausius relation holds for all the local Rindler causal horizons
through each spacetime point. Jacobson interpreted $dQ$ as the
energy flux and $T$ as Unruh temperature seen by an accelerated
observer just inside the horizon. Jacobson's work can be stated as follows; from a thermodynamic point of view, Einstein
field equations are nothing but an equation of state for the spacetime under consideration. Inspired by Jacobson approach, Cai et.al \cite{Cai}
derived Friedman equations of $(n+1)$-dimensional
Friedman-Robertson-Walker (FRW) universe, from the Clausius
relation $(TdS=-dE)$ with the apparent horizon of FRW universe,
assuming that the entropy is proportional to the area of apparent
horizon \cite{Cai}. The derivation of Cai et al. was based on
assuming that the apparent horizon has an entropy $S$, and a temperature $T$;

\bea S&=& \frac{A}{4G},\label{entropy}\\
T&=& \frac{1}{2\pi \tilde{r}_A}~ \label{Hawking} \eea
Notice here that the entropy is proportional to the area, but the temperature is not proportional to the surface gravity $\kappa$, since
\be
\kappa=- \frac{1}{\tilde{r}_A}\le(1-\frac{\dot{\tilde{r}}_A}{2~H~\tilde{r}_A}\ri).
\ee
As a result, the first law in this context is an approximation, where $\frac{\dot{\tilde{r}}_A}{2~H~\tilde{r}_A}\mid_{_{horizon}}={\dot{H} \over 2H^2}<<1$, has to be satisfied. This has been noticed and discussed by the authors in \cite{Cai}. The drawback of this approximation is that it constrains possible equations of state (EoS) to be $p\simeq -\rho$, i.e., EoS has to be close enough to that of the vacuum energy or de Sitter spacetime. This can be shown clearly through the cosmological equation $\dot{H}=-{3H^2 \over 2}\,(p/\rho+1)$ for a non-vanishing $H$. To improve this thermodynamic description of FRW cosmology, Akbar and Cai\cite{Akbar:2006kj} used the Misner-Sharp energy relation inside a
sphere of radius $\tilde{r}_A$ of an apparent horizon and rewrite the dynamical Friedman equation in the form
of the first law of thermodynamics with a work term.

\bea dE= T~dS+ W~dV, \eea where the work density $W$ is given in terms of the energy density $\rho$ and the pressure of matter in the universe $p$
as follows:

\bea
 W=\frac{1}{2}\le(\rho-p\ri).
\eea
In addition, they considered the Misner-Sharp energy to be the total energy of the matter existing
inside the apparent horizon which is given by $E=\rho V$, where $V$
is the volume of the apparent horizon.

Motivated again, by black hole physics, there is an another interesting connection between the
Hawking radiation and the uncertainty principle
\cite{Eliasentropy,Adler,Cavaglia:2003qk,Majumder} where the black
hole can be modeled as $n$-dimensional sphere of size equal to
twice of Schwarzschild radius. Since the Hawking radiation is a
quantum process, one could assume that the emitted particles should obey
the Heisenberg uncertainty relation. From this, one can derive exactly the
Hawking temperature and the thermodynamic properties of the black
hole \cite{Adler,Cavaglia:2003qk}. From the black hole thermodynamics, it has been realized
that; as the black hole approaches zero mass, its temperature
approaches infinity with infinite radiation rate which is
considered a \emph{catastrophic} evaporation of the black hole.

A generalized uncertainty principle(GUP) was proposed by
different approaches to quantum gravity such as string theory and
black hole physics\cite{guppapers,BHGUP,Scardigli,kmm,kempf,brau} in which Planck length plays an important role.
The existence of Planck length as a minimal observable
length $l_p$ is a universal feature among all approaches of QG
\cite{guppapers,BHGUP,Scardigli}. This minimal length works as a natural
cutoff, which is expected to have a crucial role in resolving curvature singularities in
general relativity. By employing this GUP, it was found that the black hole thermodynamics quantities, such as temperature, and entropy are completely modified such that the end-point of Hawking radiation is
not catastrophic anymore. This is because the GUP implies the existence of
black hole \emph{remnants} at which the specific heat vanishes
and, therefore, the black hole cannot exchange heat with the
surrounding space \cite{Eliasentropy,Adler,Cavaglia:2003qk,Majumder}. This means that GUP prevents the black hole
from evaporating completely, just like the standard  uncertainty
principle prevents the hydrogen atom from collapsing. It is worth
mentioning that the GUP modifies, significantly, the entropy-area law
as has been pointed out by several authors (see for example Ref.'s \cite{Eliasentropy,Majumder}).

In this work we study exact modifications of the entropy-area law due to GUP, which is used to modifying Friedmann equations derived by Akbar and Cai in \cite{Akbar:2006kj} at very high densities. It is worth mentioning that GUP has been studied with Friedmann equations through different frameworks in \cite{Majumder:2011eg,Lidsey:2009xz}.
In this paper, we first derive the modified Friedmann equations for
a general form of the entropy as a function of area using the
first law of thermodynamics $(dE=T~dS+W~dV)$. This generalization can host all possible
correction to the entropy-area law like the logarithmic correction
and power law corrections which are motivated by different
approaches to quantum gravity such as string theory and loop
quantum gravity\cite{log}. We then investigate an important entropy-area law which is motivated by the GUP and study the modified Friedmann equations derived through the first law approach introduced by Akbar and Cai in \cite{Akbar:2006kj}. Studying the resulted Friedmann equations reveals the existence of a maximum energy density closed to Planck density. Assuming a general continuous pressure $p(\rho,a)$ leads to bounded curvature invariants and a general nonsingular evolution. As a result, the maximum energy density is reached in a finite time and there is no cosmological evolution beyond this point from a spacetime prospective. The existence of maximum energy density and a general nonsingular evolution is independent of the equation of state and the spacial curvature $k$. As an example we study the evolution of the equation of state $p=w\rho$, using a phase-space diagram\cite{Adel}, to show the existence of a maximum energy density and a finite time to reach it. Our results reveal that the big bang singularity is not accessible in this description since the spacetime itself can not be extended beyond Planck density as a result of the GUP and the thermodynamic approach to gravity which modifies Friedmann equations.

The paper is organized as follows, in Sec.~\ref{AH}, we review the
derivation of Friedmann equations from the first law of thermodynamics due to Akbar and Cai \cite{Akbar:2006kj}, assuming that the
entropy is proportional to the area of apparent horizon. In Sec.~\ref{general}, we extend this procedure for an arbitrary entropy, which could host various possible correction to entropy-area law, to obtain a set of modified Friedmann equations.
In Sec.~\ref{GUP}, we review the generalized uncertainty principle and calculate the exact entropy-area law from the first law of thermodynamics. In
Sec.~\ref{FRW-GUP}, we calculate the modified Friedmann equations due to the exact entropy-area law obtained from GUP using the first law of thermodynamics. In Sec.~\ref{RayChud}, we discuss direct implications of the modified Friedmann equations arguing for the existence of a maximum energy density closed to Planck density. Assuming a continuous pressure we show that all curvature invariants are finite and the previous two features are independent of the equation of state and the spacial curvature. As an example we study the evolution of the equation of state $p=w\rho$, using a phase-space diagram, to show the existence of a maximum energy density and a finite time to reach it. At the end we conclude by showing the general implications of the modified Friedmann equations that we found due to the exact entropy-area law obtained from the GUP.

\section{Friedmann Equations from the first law of thermodynamics}
\la{AH}

In this section, we review the derivation of Friedmann equations
from the first law of thermodynamics relation with the apparent horizon of FRW
universe with assuming that the entropy is proportional to the
area of the apparent horizon \cite{Jacobson,Cai}. The $(n+1)$-dimensional
Friedmann-Robertson-Walker (FRW) universe is described by the
following metric:
\be
ds^2={h}_{a b}dx^{a} dx^{b}+\tilde{r}^2
d\Omega^2_{n-1}, \label{metric}
\ee
where $\tilde{r}= a(t) r,~ x^a=(t,r), h_{a b}=(-1,
a^2/(1-kr^2))$, $d\Omega^2_{n-1} $ is the metric of $(n-1)$-dimensional sphere,
$a,b=0,1$ and the spatial curvature constant $k$ takes the values
$0,1,-1$ for a flat, closed and open universe, respectively. The
dynamical apparent horizon is determined by the relation
$h^{ab}\partial_{a}\tilde{r}\partial_{b}\tilde{r}=0$, which would
give the radius of the apparent horizon to be\cite{Cai}:

\be
\label{app-radius}
\tilde{r}_A=a ~r=\frac{1}{\sqrt{H^2+k/a^2}},
\ee
where $H=\dot{a}/a$ is the Hubble parameter. By assuming that
the matter which occupy the FRW universe forms a perfect fluid, so
the energy-momentum tensor would be:
\be
T_{\mu\nu}=(\rho+p)u_{\mu}u_{\nu}+pg_{\mu\nu}.\la{stress-tensor}
\ee
where $u_{\mu}$ is the four velocity of the fluid.
The energy conservation law ( $T^{\mu\nu}_{~~~;\nu}=0$) leads to the continuity
equation

\be
\dot{\rho}+n H(\rho+p)=0.\label{Continuity}
\ee
Based on the arguments of \cite{Hayward}, one can define work
density $W$ as follows

\bea
W=-\frac{1}{2} T^{ab} h_{ab} ,
\eea
where $T_{ab}$ is the projection of the energy-momentum tensor
$T_{\mu\nu}$ in the normal direction. For FRW universe with
perfect fluid, the work density will be

\bea
W&=&\frac{1}{2} (\rho-p).
\eea
Now, we give a brief review for the procedure that has been followed by Akbar-Cai in \cite{Akbar:2006kj}.
Considering the first law of thermodynamics as follows:

\bea
dE= T~dS+ W~dV. \label{1st}
\eea
Let us calculate term by term in Eq. (\ref{1st}). We start with $dE$.
The term $dE$ represents the infinitesimal change in the total energy during
small interval of time $dt$. Since $E$ introduces the total energy of matter inside
the apparent horizon (Misner-Sharp energy), so it can take the following form
\bea
E=\rho~ V ,
\eea
where $V=\Omega_n \tilde{r}_A^n$ is the volume of n-dimensional sphere with radius $\tilde{r}_A$
and $\Omega_n=\frac{\pi^{n/2}}{\Gamma(n/2+1)}$. Hence, the differential element of $E$ would be
\bea
dE&=& \rho dV+ V~ d\rho \nn\\
&=& n\Omega_n\tilde{r}_A^{n-1} \rho d\tilde{r}_A+\Omega_n\tilde{r}_A^n d\rho.\label{denergy}
\eea
By using the continuity equation of Eq. (\ref{Continuity}) in Eq. (\ref{denergy}), we get;
\bea
dE= n\Omega_n\tilde{r}_A^{n-1} \rho d\tilde{r}_A- n \Omega_n \tilde{r}_A^n (\rho+p) H dt \label{denergy1}
\eea
Turning to the other term $W~dV$, it can be written as follows:
\bea
W~dV= \frac{1}{2} n \Omega_n \tilde{r}_A^{n-1} \le(\rho-p\ri)d\tilde{r}_A. \label{dwork}
\eea
For the term $T ~ dS$, we should use definition of Hawking temperature of Eq. (\ref{Hawking}) as well as the entropy-area law.
We start with Hawking temperature which is defined in terms of surface gravity as follows:

\bea
T= \frac{\kappa}{2\pi}
\eea
where $\kappa$ introduces the surface gravity for the metric of Eq. (\ref{metric}) and is defined as

\bea
\kappa&=&\frac{1}{2\sqrt{-h}}\partial_a(\sqrt{-h}h^{ab}\partial_b\tilde{r}) \nn\\
&=&- \frac{1}{\tilde{r}_A}\le(1-\frac{\dot{\tilde{r}}_A}{2~H~\tilde{r}_A}\ri)
\eea
Besides, we use the entropy-area law of Eq. (\ref{entropy}) and the expression of area
for n-dimensional sphere $A=n \Omega_n \tilde{r}_A^{n-1}$. This yields at the end the following expression

\bea
T~dS&=& \frac{\kappa}{2\pi} d\le(\frac{n\Omega_n \tilde{r}_A^{n-1}}{4G}\ri)\nn\\
&=& -\frac{1}{2\pi\tilde{r}_A} \le[1-\frac{\dot{\tilde{r}}_A}{2~H~\tilde{r}_A}\ri] \le(\frac{n(n-1)\Omega_n}{4G}\tilde{r}_A^{n-2}\ri) \label{TdS}
\eea
By substituting Eqs. (\ref{denergy1}, \ref{dwork} and \ref{TdS}) into the first law of thermodynamics of Eq. (\ref{1st}), we get
\bea
\frac{d\tilde{r}_A}{\tilde{r}_A^3}= \frac{8\pi G}{n-1} \le(\rho+p\ri)~H dt. \label{1frw}
\eea
Using Eq. (\ref{app-radius}) which yields $d\tilde{r}_A= -H \tilde{r}_A^3\le(\dot{H}-k/a^2\ri) dt$ \cite{Cai}, one simply finds that
 Eq. (\ref{1frw}) introduces the dynamical Friedman equation.

\bea \dot{H}-\frac{k}{a^2}=-\frac{8 \pi G}{n-1}(\rho+p)
\label{dotH} \eea Using the continuity equation (\ref{Continuity})
and integrating Eq. (\ref{dotH}), one simply gets \bea
H^2+\frac{k}{a^2}=\frac{8\pi G}{n(n-1)}\rho, \eea which is the
Friedmann equation for $(n+1)$-dimensional FRW universe.

\section{Modified Friedmann Equation for a general form of the entropy}
\label{general}

In this section, we derive the general
modified Friedmann equation for a general expression of the
entropy as a function of area  through the apparent horizon approach
and first law of thermodynamics. Suppose that the general expression for the entropy-area relation
takes the following form: \bea
S&=& \frac{f(A)}{4G}, \\
\frac{dS}{dA}&=& \frac{f^{\prime}(A)}{4G},
\eea
where $f^{\prime}(A)=df(A)/dA$.
Using the first law of thermodynamics  $dE= T~dS+W~dV$, and following the same procedure
that we reviewed in Sec. \ref{AH}, we get

\bea
f^{\prime}(A) \frac{d\tilde{r}_A}{\tilde{r}_A^3}= \frac{8 \pi G}{n-1} \le(\rho+p\ri)~H~dt \label{dr}
\eea
Again, using Eq.(\ref{app-radius}) which yields\cite{Cai},
\bea
d\tilde{r}_A= -H \tilde{r}_A^3\le(\dot{H}-k/a^2\ri) dt
\eea
one simply finds that Eq. (\ref{1frw}) introduces the dynamical Friedman equation.
\bea
\le(\dot{H}-\frac{k}{a^2}\ri) f^{\prime}(A)= -\frac{8 \pi G}{n-1} \le(\rho-p\ri) \label{FR1}
\eea
Using the continuity equation of Eq. (\ref{Continuity}), and with few calculations, we can rearrange Eq. (\ref{dr}) as follows
\bea
f^{\prime}(A) \frac{(n\Omega_n)^{\frac{n+1}{n-1}}}{n(n-1)\Omega_n} \frac{dA}{A^{\frac{n+1}{n-1}}}= -\frac{8\pi G}{n(n-1)} d\rho
\eea
The last equation can be integrated to give the general modified Friedmann equation for $(n+1)$-dimensional FRW universe due to
a general form of the entropy using the first law of thermodynamics.
\bea
-\frac{8 \pi G}{n(n-1)} \rho= \int \frac{(n\Omega_n)^{\frac{n+1}{n-1}}}{n(n-1)\Omega_n} f^{\prime}(A) \frac{dA}{A^{\frac{n+1}{n-1}}}  \label{FR2}
\eea
Again, if we set $f(A)=A$, the first Friedmann equation will be satisfied. These
modified equations was gotten first in \cite{Cai:2008ys} using the Clausius relation $dE= T~dS$. We are deriving them here
with considering the contribution of the work $W$ in the first law of thermodynamics.

\section{The Generalized Uncertainty Principle and Entropy-Area Law}
\label{GUP}
We first review briefly the Generalized uncertainty
principle (GUP) \cite{guppapers,BHGUP} and secondly we review its
effect on the area-entropy law
\cite{Eliasentropy,Majumder,Adler,Cavaglia:2003qk}. The existence of a minimum
measurable length originates as an intriguing prediction of various frameworks of
quantum gravity such as string theory \cite{guppapers} and black hole
physics \cite{BHGUP}. This implies a direct modification of the
standard uncertainty principle\cite{guppapers,BHGUP,Scardigli,kmm,kempf,brau}:

\bea \Delta x \gtrsim \frac{\hbar}{\Delta p}\left[1+ \frac{\beta~
\ell_{P}^2}{\hbar^2} (\Delta p)^2\right], \label{GUP} \eea where
$\ell_{P}$ is the Planck length and $\beta$ is a dimensionless
constant which depends on the quantum gravity theory. The new
correction term in Eq. (\ref{GUP}) becomes effective when the
momentum and length scales are of order the Planck mass and of the
Planck length, respectively. It was straightforward to find that Eq.(\ref{GUP})
implies the existence of minimal measurable length scale as follows:

\bea
\Delta x \gtrsim \Delta x_{min} = 2 \beta~ \ell_{P}
\eea
By some manipulations, GUP can be represented by another
form as follows:
\bea
\frac{\Delta p}{\hbar} \gtrsim \frac{\Delta x}{2 \beta~ \ell_P^2}\le[1- \sqrt{1-\frac{4 \beta~ \ell_P^2}{\Delta x^2}}~\ri]
\eea
There has been investigations devoted to study the impact of GUP on
the black hole thermodynamics and to the Bekenstein--Hawking
(black hole) entropy (e.g.,
\cite{Eliasentropy,Majumder,Adler,Cavaglia:2003qk}). These studies
are based on the argument that Hawking radiation is a quantum
process and it should respect the uncertainty principle.

Here we review the analysis of \cite{Eliasentropy} and use the arguments in
\cite{AmelinoCamelia:2004xx} which say that the uncertainty
principle $\D p \geq 1/ \D x$  can be represented by the lower
bound $E\geq 1/ \D x$, so one can get for the GUP case:

\bea
E \gtrsim \frac{\Delta x}{2 \beta \ell_P^2} \le[1- \sqrt{1-\frac{4 \beta~ \ell_P^2}{\Delta x^2}}~\ri]. \label{energy0}
\eea
For any black hole absorbing or emitting a quantum particle whose energy $E$ and
size $R$, the area of the black hole would change by an
amount \cite{Areachange}.

\be \Delta A \geq 8 \pi\, \ell_p^2\, E\, R, \ee The quantum
particle itself implies the existence of finite bound given by \be
\Delta A_{min} \geq 8 \pi\, \ell_p^2\, E\, \Delta\, x. \la{Darea}
\ee
Using GUP by  substituting  (\ref{energy0}) into (\ref{Darea}), we get
\bea
\Delta A_{min} \gtrsim \frac{8 \pi \Delta x^2}{2 \beta} \le[1- \sqrt{1-\frac{4 \beta~ \ell_P^2}{\Delta x^2}}~\ri]. \label{Area}
\eea
The value of $\Delta x$  in this analysis is set to be the inverse of surface gravity
$\D x= \kappa^{-1}= 2 r_s$ where $r_s$ is the Schwarzschild
radius, where this is  probably the most sensible choice of length
scale in the context of near-horizon geometry
\cite{Eliasentropy,Adler,Cavaglia:2003qk}. This implies the
following identity: \be \D x^2 = \frac{A}{\pi}. \label{DX} \ee
Substituting  Eq. (\ref{DX}) into Eq. (\ref{Area}), we get
\bea
\Delta A_{min} \gtrsim \frac{8~A}{2 \beta}\le[1- \sqrt{1-\frac{4 \beta~ \ell_P^2 \pi}{A}}~\ri].
\eea
The area change is then determined as:
\bea
\Delta A_{min} \simeq \lambda \frac{8~A}{2 \beta}\le[1- \sqrt{1-\frac{4 \beta~ \ell_P^2 \pi}{A}}~\ri],
\eea
where the parameter $\lambda$ will be fixed from the
Bekenstein-Hawking entropy formula. According to
\cite{Bekenstein,Hawking, Bardeen}, the black hole's entropy is
conjectured to depend on the horizon's area. From the information
theory \cite{Adami:2004mx}, it has been stated that the minimal
increase of entropy should be independent of the area. It is just
one bit of information which is $\Delta S_{min}=b = \ln(2)$.

\be
\frac{dS}{dA}= \frac{\Delta S_{min}}{\Delta A_{min}} = \frac{b}
{\lambda \frac{8~A}{2 \beta}\le[1- \sqrt{1-\frac{4 \beta~ \ell_P^2 \pi}{A}}~\ri]}.
\ee
According to \cite{Eliasentropy}, the Bekenstein-Hawking entropy
formula has been used to calibrate the the constants $b/\lambda= 2
\pi$, so the we have
\bea \frac{dS}{dA}= \frac{\Delta S_{min}}{\Delta A_{min}} =
\frac{\pi} {\frac{2~A}{\beta}\le[1- \sqrt{1-\frac{4 \beta~
\ell_P^2 \pi}{A}}~\ri]} \label{Dentropy} \eea
To simplify the expression of Eq. (\ref{Dentropy}), we set $\alpha= 4
\beta \ell_P^2 \pi$, so we get
\bea \frac{dS}{dA}= \frac{\alpha}{8 \ell_P^2} \frac{1}{A \le[1-
\sqrt{1-\frac{\alpha}{A}}~\ri]}. \label{exact} \eea

In this paper, we are interested in the \emph{exact} form of the
entropy instead of the approximated one that was used in
\cite{Eliasentropy}, so we will not make any approximation for the
expression of $dS/dA$. The exact form of Eq. (\ref{exact})
would enable us to study the behavior of a general solution of the
modified Friedmann equations (using its fixed points) due to GUP
through the first law of thermodynamics. To get the exact expression of the entropy, we integrate Eq.
(\ref{exact}) to yield:

\bea
S= &&\frac{1}{8 \ell_P^2} \Bigg[
A+\sqrt {{A}^{2}-A\alpha}-\frac{\alpha}{2}\,\ln  \left(
A+\sqrt {{A}^{2}-A\alpha}-\frac{\alpha}{2} \right)\Bigg]+ S_0
\label{exactentropy}
\eea
where $S_0$ is an integration constant.
We find that Eq. (\ref{exactentropy})
modifies Bekenstein-Hawking entropy as a result of a minimum length scale or the GUP considered above. In the next section,
we will calculate the modified Friedmann equations due to the
modified entropy, Eq. (\ref{exactentropy}), of the apparent
horizon of FRW universe.

\section{Modified Friedmann Equations due to GUP}
\label{FRW-GUP}

In this section, we implement the modified entropy of
Eq.(\ref{exactentropy}) in the first law of thermodynamics relation $dE=T dS+W dV$, and
derive the modified Friedmann equations with the apparent horizon
approach\cite{Jacobson,Cai,Akbar:2006kj}. Now, we consider our current
case of modified entropy due to GUP
in Eq. (\ref{exact}) and Eq. (\ref{exactentropy}). By substituting
Eq. (\ref{exact}) into the general modified Friedmann equations
Eq. (\ref{FR1}) and Eq. (\ref{FR2}) with $n=3$, we get the following
expression:

\bea
\le(\dot{H}-\frac{k}{a^2}\ri) \,\frac{\alpha}{2}\, \frac{1}{A-\sqrt{A^2-A\,\alpha}}=
-4 \pi G (\rho+p), \label{MFR1}
\eea
\bea
\frac{8 \pi G}{3} \rho &=& - 16 \pi \ell_P^2 \int \frac{1}{A^2}\, \frac{\alpha}{8 \ell_P^2}\, \frac{1}{A-\sqrt{A^2-\alpha\,A}} dA, \nn\\
&=& 2 \pi \le(\frac{1}{A}-\frac{2(A^2-\alpha A)^{3/2}}{3 \alpha A^3}\ri)+C,\label{MFR2}
\eea
where $C$ is a constant of integration and it can be fixed from the initial conditions in Eq. (\ref{MFR2}).
As the universe expands, the area of apparent horizon is supposed to go to infinity with the density having vacuum energy density $\rho_{\text{vac}}=\Lambda$, where
$\Lambda$ is the cosmological constant.
From this argument, $C$ takes the following value:
\bea
C= \frac{8 \pi G}{3} \Lambda + \frac{4 \pi}{3 \alpha} = \frac{8 \pi G}{3} \le(\Lambda+ \frac{1}{2 G \alpha}\ri).
\eea
The area of the apparent horizon is given in Sec. (\ref{AH}) as: \cite{Cai}

\bea
A= 4 \pi \tilde{r}_A^2 = \frac{4 \pi}{H^2+\frac{k}{a^2}}.
\eea
Accordingly, the modified Friedmann equations (\ref{MFR1}) and
(\ref{MFR2}) due to GUP will be

\bea
\frac{8 \pi G}{3}(\rho-\Lambda) &=& \frac{1}{2} \le(H^2+\frac{k}{a^2}\ri)+{4\,\pi \over 3 \alpha}\,\left[1-\le(1-\frac{\alpha}{4\, \pi}\,\le(H^2+\frac{k}{a^2}\ri)\ri)^{\frac{3}{2}}\right], \label{MFRA}\\
- 4 \pi G(\rho+p)&=&\le(\dot{H}-\frac{k}{a^2}\ri) \frac{\alpha}{8\, \pi}\frac{(H^2+\frac{k}{a^2})} {\le[1-\le(1-\frac{\alpha}{4\,\pi}\le(H^2+\frac{k}{a^2}\ri)\ri)^{\frac{1}{2}}\ri]}.\label{MFRB}
\eea
We find that Eq. (\ref{MFRA}) and Eq. (\ref{MFRB}) give the exact
modified Friedmann equations due to GUP using Akbar-Cai approach
\cite{Akbar:2006kj}. In the next section we are going to discuss consequences of Eq. (\ref{MFRA}) and Eq. (\ref{MFRB}) on the behavior of the FRW cosmology.

\section{Maximum Density and Curvature Singularities}
\label{RayChud}
It is interesting to notice that Eq. (\ref{MFRA}) leads to a bounded energy density $\rho$ since the inequality $H^2+{k \over a^2} \leq 4 \pi/\alpha$ must be satisfied, otherwise, the density is complex. Let us analyze this more closely for cases with $k=0$, and $k=1$. In order to keep the density real the previous inequality requires "a" to have a minimum and "H" to have a maximum. This leads to a maximum energy density $\rho_{max}=\Lambda+{5 \over 4G\alpha}\sim \rho_p$ (since $\Lambda$ is tiny), where $\rho_p$ is Planck's density.

Having a maximum value for $H$ (or $\rho$ since they are related) does not necessarily means a finite curvature or nonsingular behavior. The reason is that curvature invariants such as Ricci scalar, Riemann tensor squared, etc., depend not only on $H$ but also on $\dot{H}$. But the behavior of $\dot{H}$ is controlled by both $\rho$ and $p$ as it is clear from Eq. (\ref{MFRB}). To check whether $p$ has a diverging behavior or not we don't have to know the equation of state, it is enough to know some generally properties of pressure. Here we will follow the discussion in Ref. \cite{Adel} on general properties of pressure which leaves the mathematical problem addressed by Eq. (\ref{MFRA}) well defined. Here we assume that the pressure, $p(H,a)$ or $p(\rho,a)$ is a continues of function of its arguments, in order to have a well defined mathematical evolution for $H$ or $\rho$. This ensures existence and uniqueness of the general solution of the continuity equation or Eq. (\ref{MFRB}). In addition, having a discontinues pressure in $H$ or $\rho$ could imply a noncausal evolution since it leads to a divergent $dp/d\rho$, or an unbounded speed of sound! Therefore, it is natural to assume a continues $p(H,a)$ or $p(\rho,a)$. As a result, one observes that if $a$ has a minimum value and $H$ is bounded, then $\dot{H}$ must be bounded given Eq. (\ref{MFRB}). Therefore, all curvature invariants of the Friedmann-Robertson-Walker metric are finite since they can be expressed as functions of $H$ and $\dot{H}$.

 But what about the $k=-1$ case, which potentially could still leads to a singular solution. Here we need to satisfy the inequality $H^2+{k \over a^2} \leq 4 \pi/\alpha$, for this case too. For a general continues pressure $p(H,a)$, such that $H^2 \neq a^{-2}$ as $a\rightarrow 0$, the previous inequality leads to a minimum value for $a$  and maximum value for $H$. But for the specific case where $H^2 = a^{-2} + c_1$ as $a\rightarrow 0$ the scale factor $a$ can go to zero and the Hubble rate to infinity without violating the inequality. In this case, it is interesting to notice that both density and  pressure are finite. A finite pressure can be shown through taking the derivative of $H$ with respect to time using $H = \sqrt{a^{-2} + c_1}$ and calculating $p$ using Eq. (\ref{MFRB}). Analyzing this case we found that it leads to a specific EoS, namely $p=-\rho$, which gives a de Sitter evolution as $a\rightarrow 0$.
Our conclusion is that the above modified Friedmann equation leads to a bounded energy density with a maximum value given by the Planck's density for any equation of state and all values $k$. In addition, by assuming a continues pressure $p(\rho,a)$ one can show that curvatures invariants are finite as a result of maximum density $\rho$ and minimum scale factor $a$, therefore, the solution is nonsingular. Our results reveals the limitation or the breaking down of the spacetime description of gravity near Planck scale energies which was not evident from general relativity. As an example we will discuss next the EoS $p=\omega \rho$ and its Raychaudhuri equation which describes a nonsingular universe which starts from a finite time with a Planck density.

Now to show some of the features discussed above let us consider the usual equation of state $p= \omega \rho$, in the light of this modified Friedmann equation and check the behavior of the universe in early times. To do that one can study Raychaudhuri equation or Eq. (\ref{MFRB}) on the following form\cite{Adel}

\be
\dot{H}= F(H) \label{raych}
\ee

Using the modified Friedmann equations of Eq. (\ref{MFRA}) and Eq. (\ref{MFRB}), and the above equation of state, we get the following form of Raychaudhuri equation.

\bea
\le(\dot{H}-\frac{k}{a^2}\ri)&&=-\frac{3}{2} (1+\omega)\Bigg[\frac{1}{2}\le(H^2+\frac{k}{a^2}\ri)-\frac{4 \pi}{3 \alpha}\le(1-\frac{\alpha}{4 \pi}\le(H^2+\frac{k}{a^2}\ri)\ri)^{\frac{3}{2}}+C\Bigg] \times \nn\\ &&
 \le(\frac{8\, \pi}{\alpha}\frac{\le(1-\le(1-\frac{\alpha}{4 \pi} \le(H^2+\frac{k}{a^2}\ri) \ri)^{\frac{1}{2}}\ri)}{\le(H^2+\frac{k}{a^2}\ri)} \ri).
\eea

For the flat case with $k=0$ the Raychaudhuri equation takes the following form

\bea
\dot{H}= &&-\frac{3}{2} (1+\omega) \le[\frac{H^2}{2}-\frac{4 \pi}{3 \alpha} \le(1-\frac{\alpha}{4\pi} H^2\ri)^{\frac{3}{2}}+\frac{8 \pi G}{3} \Lambda + \frac{4 \pi}{3 \alpha} \ri]\nonumber\\
&& \times \le[\frac{8\, \pi}{\alpha}\frac{\le(1-\le(1-\frac{\alpha}{4 \pi} H^2 \ri)^{\frac{1}{2}}\ri)}{H^2} \ri] \label{mf}
\eea

\begin{figure}[htb!]
\centering
\includegraphics[scale=0.4,width=6.cm,angle=270]{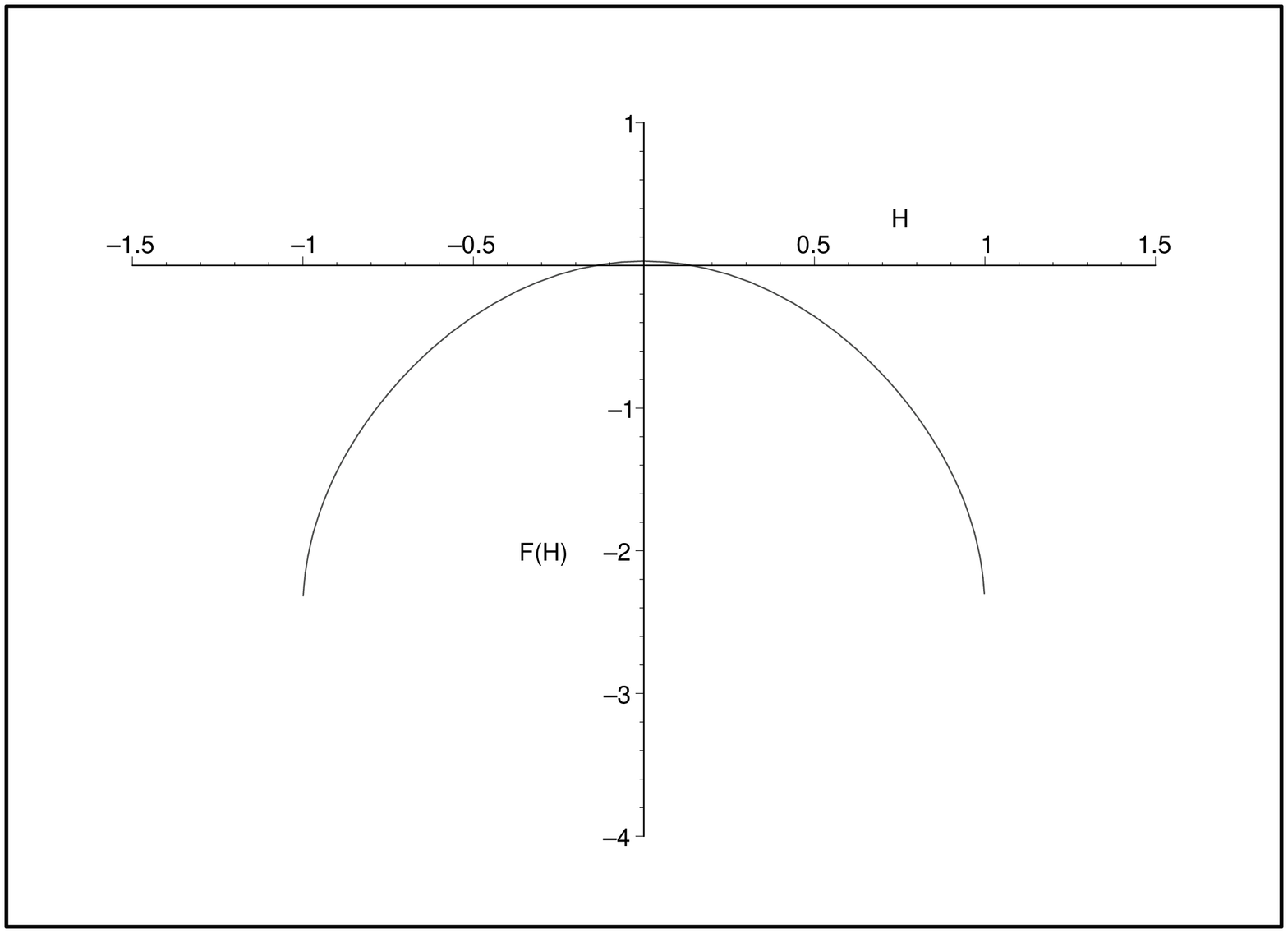}
\caption{$\dot{H}=F(H)$ versus $H$, $4\pi/\alpha =1$}. $H$ has a maximum value $H_{max}\sim \sqrt{\rho_p/3}$.
\label{solution}
\end{figure}

Using the general analysis of the Eq. (\ref{raych}) in \cite{Adel} for FRW cosmology. The above first-order system is
well studied in dynamical system in cosmological context, see e.g., \cite{Adel}, or see \cite{strogatz} for more general
applications. Knowing $F(H)$ fixed points (i.e., its zeros. Let us
call them $H_i$) and its asymptotic behavior enables one to
qualitatively describe the behavior of the solution without
actually solving the system \cite{Adel}.
Drawing the phase-space diagram for For Eq. (\ref{mf}) for $\omega=1/3$ (i.e., plotting $\dot{H}$ versus $H$) we get the following behavior in
Fig.(\ref{solution})

As one can notice we have a future fixed point at $H_f=\sqrt{\Lambda/3}$, which we can provide the asymptotic behavior of our universe at late times. In addition this fixed point is reached in an infinite time since by integrating Eq. (\ref{mf}) we get
\be t= \int_{H_0}^{H_f}\,{ dH \over F(H)} =\infty, \ee
where $H_0$ is the initial value of the Hubble rate and F(H) is
\bea F(H)=-\frac{24 \pi(1+\omega)}{2\alpha H^2} \le[\frac{H^2}{2}-\frac{4 \pi}{3 \alpha} \le(1-\frac{\alpha}{4\pi} H^2\ri)^{\frac{3}{2}}+\frac{8 \pi G}{3} \Lambda + \frac{4 \pi}{3 \alpha} \ri] \le(1-\le(1-\frac{\alpha}{4 \pi} H^2 \ri)^{\frac{1}{2}}\ri)\eea
The interesting observation in this diagram is the existence of a maximum value for the Hubble rate $H$, beyond which there is no evolution allowed. We have calculated the time to reach the maximum Hubble rate $H_{max}$
\be t_m= -\int_{H_0}^{H_{max}}\,{ dH \over F(H)}= finite ,\ee
 Since all curvature invariants of the FRW metric are functions of the Hubble rate and it first-time-derivative, it is straight forward to show that they are all finite as a result of a maximum density and the EoS $p=w\rho$ given the above modified Friedmann equations. It is interesting to observe that the big bang singularity is not accessible in this description since the spacetime itself can not be extended beyond Planck density as a result of the minimum length or the GUP and the thermodynamic approach to gravity provided which modifies Friedmann equations.

\section{Conclusions}
We generalize Akbar--Cai derivation \cite{Akbar:2006kj} of Friedmann equations from the first law of thermodynamics $dE=T dS+ W dV$, to include an arbitrary entropy-area law which could include possible corrections arise from different approaches to quantum gravity. Studying the resulted Friedmann equations for an entropy-area law motivated by the generalized uncertainty principle (GUP) revealed the existence of a maximum energy density with a value around Planck density. Allowing for a general continuous pressure $p(\rho,a)$ lead to bounded curvature invariants and a general nonsingular evolution. In this case, the maximum energy density is reached in a finite time and there is no cosmological evolution beyond this point from a spacetime prospective. The existence of maximum energy density and a general nonsingular evolution is independent of the equation of state and the spacial curvature $k$. As an example we study the evolution of the equation of state $p=w\rho$, using a phase-space diagram, to show the existence of a maximum energy density and a finite time to reach it. Our results reveal that the big bang singularity is not accessible in this description since the spacetime itself can not be extended beyond Planck density as a result of the minimum length or the GUP and the thermodynamic approach to gravity which modifies Friedmann equations.

\subsection*{Acknowledgments}
The research of AFA is supported by Benha University (www.bu.edu.eg) and CFP in Zewail City.



\end{document}